\documentclass[12pt]{article}

\title{\bf Signature for the Shape of the Universe}
\author{G.I. Gomero\thanks{E-mail: german@cbpf.br} , \  
M.J. Rebou\c{c}as\thanks{E-mail: reboucas@cbpf.br} , \
 and A.F.F. Teixeira\thanks{E-mail: teixeira@cbpf.br} \\ 
\\ 
Centro Brasileiro de Pesquisas F\'\i sicas, \\
Rua Dr. Xavier Sigaud 150, \\
22290-180 Rio de Janeiro -- RJ, Brazil
}

\begin{document}

\maketitle

\begin{abstract}
If the universe has a nontrivial shape (topology) the sky may show 
multiple correlated images of cosmic objects. These correlations 
can be couched in terms of distance correlations. We propose a 
statistical quantity which can be used to reveal the topological 
signature of any Robertson-Walker (RW) spacetime with nontrivial topology. 
We also show through computer-aided simulations how one can extract 
the topological signatures of  f\/lat, elliptic, and  hyperbolic
RW universes with nontrivial topology.
\end{abstract}

\section{Introduction}
 \label{intro} 

Whether we live in a f\/inite or inf\/inite space and what is the 
size and the shape of the universe are open problems, which 
modern cosmology seeks to solve (see, for example, 
\cite{LaLu}~--~\cite{GRT99a} and references therein).
The primary consequence of multiply-connectedness of the universe 
is the possibility of observing multiple images of cosmic objects, 
whose existence can be perceived by the simple reasoning 
presented below. 

In the general relativity approach to cosmological modeling
the Robertson-Walker (RW) spacetime manifolds $\mathcal{M}_4$ are 
decomposable into $\mathcal{M}_4 = \mathcal{R} \times M$, and 
endowed with the metric
\begin{equation}
\label{RWmetric}
ds^2 = dt^2 - R^2(t)\,\{\, d\chi^2 + f^2(\chi)\,[\,d\theta^2 
                         +\sin^2\theta\,d\phi^2\,] \,\} \;,
\end{equation}
where $t$ is a cosmic time, the function $f(\chi)$ is given by 
$f(\chi)= \chi\,,\;$ $\sin\chi\,,\;$ or $\sinh\chi\,,\;$ 
depending on the sign of the constant spatial curvature 
($k = 0, \pm 1$), and $R(t)$ is the scale factor. 

It is often assumed that the $t=const$ spatial sections $M$ of RW 
spacetime manifold $\mathcal{M}_4$ are one of the following simply 
connected spaces:  Euclidean $E^3$ ($k=0$), elliptic $S^3$ ($k=1$), 
or the hyperbolic $H^3$ ($k=-1$), depending on the sign of the  
curvature $k$.
However, the simply-connectedness for our 3-space has not been
settled by cosmological observations. Thus, the space $M$ where we
live may be any one of the possible multiply-connected quotient 
3-spaces $M = \widetilde{M}/\Gamma$, where $\widetilde{M}$ stands 
for $E^3$, $S^3$ or $H^3$, and $\Gamma$ is a discrete group 
of isometries acting freely on the covering simply-connected 3-space 
$\widetilde{M}$.
The action of $\Gamma$ tessellates $\widetilde{M}$ into identical 
cells or domains which are copies of what is known as fundamental 
polyhedron (FP). In forming the quotient manifolds $M$ the 
essential point is that they are obtained from $\widetilde{M}$ 
by identifying points which are equivalent under the action of 
the group $\Gamma$. Hence, each point on the quotient 
manifold $M$ represents all the equivalent points on the 
covering manifold $\widetilde{M}$.
In this cosmological modeling context a given cosmic object is 
described  by a point $p \in M$, which represents, when $M$ is 
multiply-connected, a set of equivalent points (images of $p$) 
on the covering manifold $\widetilde{M}$. 
So, to f\/igure out that multiple images of an object can indeed be 
observed if the universe is multiply-connected, consider that the 
observed universe is a ball $\mathcal{B}_{R_H} \subset \widetilde{M}$ 
whose radius $R_H$ is the particle horizon, and denote by $L$ the 
largest length of the fundamental polyhedron FP of $M$ 
($\rm{FP} \subset \widetilde{M}$). 
Thus, when $R_H >L/2$, for example, the set of (multiple) images of 
a given object that lie in $\mathcal{B}_{R_H}$ can in principle 
be observed. Obviously the observable images of an object constitute 
a f\/inite subset of the set of all equivalent images of the object.

The multiple images of the cosmic objects are periodically distributed, 
and the periodicity, which arises from the correlations in their 
positions dictated by the group $\Gamma$, are fundamentally related 
to the topology of the 3-space $M$. The correlations among the images 
can be couched in terms of spatial distance correlations. 
Indeed, one may look for spatial distance correlations between cosmic 
images in multiply-connected universes by using pair separations 
histograms (PSH), which are functions $\Phi(s_i)$ that count the 
number of pair of images separated by a distance $s_i$ that lies 
in a given interval $J_i$ (say).
The embryonic expectation is certainly that the distance 
correlations manifest as topological spikes in PSH's, and that 
the spike spectrum of topological origin would be a def\/inite 
signature of the topology. 
While simulations performed for specif\/ic f\/lat manifolds 
appeared to conf\/irm the primary expectation~\cite{LeLaLu}, 
histograms subsequently generated for the Weeks and one of the 
Best hyperbolic manifolds revealed that the PSH's corresponding to
those specific 3-manifolds exhibit no spikes~\cite{LeLuUz,FagGaus}.
Concomitantly, a theoretical statistical analysis of the distance 
correlations in the PSH's was performed, and a formal proof was 
presented that the spikes of topological origin in PSH's are due
to just one type of isometry: the translations~\cite{GTRB98}.
This result explains the absence of spikes in the PSH's of 
hyperbolic manifolds, and also gives rise to the fact that
Euclidean distinct manifolds which admit the same translations
on their covering group present the same  spike spectrum of 
topological origin (hereafter called topological spike, 
for short).

Although the set of topological spikes in PSH's is not def\/inite 
topological signature and is not suf\/f\/icient for distinguishing 
even between some compact f\/lat manifolds~\cite{GRT99a}, the most 
striking evidence of multiply-connectedness in PSH's is indeed the 
presence of topological spikes, which arise only when the covering 
group $\Gamma$ contains translational isometries $g_t\,$.
The other  isometries $g$, however, manifest as rather tiny deformations 
of the expected pair separation histogram $\Phi^{sc}_{exp}(s_i)$ 
corresponding to the underlying simply connected universe%
~\cite{GTRB98}. 

In computer-aided simulations the histograms contain statistical 
f\/luctuations (noises), which can give rise to sharp 
peaks of statistical origin, or can hide (or mask) the tiny 
deformations due to non-translational isometries. 
The most immediate approach to cope with f\/luctuation problems 
in PSH's  is by using the mean pair separation histogram (MPSH)
scheme to obtain $<\!\Phi(s_i)\!>$  rather than a single PSH 
$\,\Phi(s_i)\,$.
However, from computer simulations it becomes clear that
for a reasonable number of images and when there is no 
topological spike (no translation) the graphs of the 
expected pair separation histograms (EPSH)
$\,\Phi_{exp}(s_i)\, \simeq \,\,<\!\Phi(s_i)\!>\,$ of multiply-%
connected universes are practically indistinguishable from the 
graph of the EPSH $\,\,\Phi^{sc}_{exp}(s_i)\,$ corresponding to 
the underlying simply connected universe, making clear that 
even the noiseless quantity $\,\Phi_{exp}(s_i)$ (PSH without 
the statistical noise) is not a suitable quantity 
for revealing the topology of multiply-connected universes.

In this work we propose a way of extracting the topological signature 
of any multiply-connected universe of constant curvature by using a 
suitable \emph{new} statistical quantity $\varphi^S(s_i)$.
We also show through computer-aided simulations the strength of our 
proposal by extracting  the topological signatures of a f\/lat ($k=0$), 
an elliptic ($k=1$), and a hyperbolic ($k=-1$) multiply-connected RW 
universe. 

\section{Topological Signature and Simulations}
\label{sig}
\setcounter{equation}{0}

Let us start by recalling that in dealing with discrete astrophysical 
sources in the context of multiply-connected RW spacetimes, the 
\emph{observable universe} is the region or part of the universal 
covering manifold $\widetilde{M}\,$  causally connected to an image 
of a given observer since the moment of matter-radiation 
decoupling. Clearly in the observable universe one has the
set of observable images of the cosmic objects, denoted by
$\mathcal{O}\,$. A catalog is a particular subset 
$\mathcal{C} \subset \mathcal{O}$, of \emph{observed} images, 
since by several observational limitations one can hardly 
record all the images present in the observable universe. 
The \emph{observed universe} is the part of the observable universe 
which contains all the sources registered in the catalog. 
Our observational limitations can be formulated through 
selection rules which dictate how the subset $\mathcal{C}$ 
arises from $\mathcal{O}$. 
Catalogs whose images obey the same well-behaved distribution
and that follow the same selection rules are said to be comparable 
catalogs. It should be noted that in the process of construction 
of catalogs it is assumed a RW geometry (needed to convert redshift 
into distance) and that a particular type of sources 
(clusters of galaxies, quasars, etc) is chosen from the 
outset. 

Consider a catalog $\mathcal{C}$ with $n$ cosmic images and 
denote by $\eta(s)$ the number of pairs of images whose spatial  
separation is $s$. Consider also that our observed universe is a 
ball of radius $a$ and divide the interval $(0,2a]$ in a number $m$ of
equal subintervals $J_i$ of length $\delta s = 2a/m$. %
\footnote{In the coordinate system relative to which the line 
element~(\ref{RWmetric}) is written this ball is def\/ined by  
$\chi \leq a\,$ for any RW metric.} 
Each of such subintervals can always be taken to be in  the form 
\begin{equation} \label{Ji}
J_i = (s_i - \frac{\delta s}{2} \,, \, s_i + \frac{\delta
s}{2}] \;, \quad
\end{equation}
with $i=1,2, \ldots ,m\,$, and centered at
\begin{displaymath}
s_i = \,(i - \frac{1}{2}) \,\, \delta s \;.
\end{displaymath}
The PSH is a normalized function which counts the number $\eta(s)$  
of pair of images separated by a spatial distance $s$ that lies in 
a given subinterval $J_i$. Thus the function PSH is given by
\begin{equation}  \label{defpsh}
\Phi(s_i)=\frac{2}{n(n-1)}\,\,\frac{1}{\delta s}\,
               \sum_{s \in J_i} \eta(s) \;, 
\end{equation}
where $n$ is the number of cosmic images in $\mathcal{C}$, and the 
PSH is clearly subjected to the normalizing condition
\begin{equation}
\sum_{i=1}^m \Phi(s_i)\,\, \delta s = 1 \; .
\end{equation}

If one considers an ensemble of comparable catalogs with the same 
number $n$ of images, and corresponding to the same 3-manifold $M$ 
of constant curvature, one can compute, e.g. probabilities, expected 
and mean values of quantities which depend on the images in the 
catalogs of the ensemble.
In particular, one can compute the expected (and normalized) pair 
separation histogram which clearly is given by 
\begin{equation}
\label{defepsh}
\Phi_{exp}(s_i)=\frac{1}{N}\,\,\frac{1}{\delta s}\,\,\eta_{exp}(s_i)
               = \frac{1}{\delta s} \,\, F(s_i) \;, 
\end{equation}
where $\eta_{exp}(s_i)$ is the number of images with separation in
the interval $J_i$, the normalizing number $N=n(n-1)/2$ is the total 
number of pairs of cosmic images in a typical catalog of the ensemble 
$\mathcal{C}$, and $F(s_i)$ is the probability that two images 
listed in $\mathcal{C}$ be separated by a distance that lies in 
$J_i = (s_i - \frac{\delta s}{2} \,, \, s_i + \frac{\delta s}{2}]\,$.
 
In multiply-connected universes there are two types of pairs, namely
$\Gamma$-pairs or correlated pairs, and $U$-pairs or uncorrelated pairs.
A $g$-pair is a pair of the form $(p, gp)$ for any (f\/ixed) isometry 
$g$.%
\footnote{When referring collectively to correlated pairs we use 
the terminology $\Gamma$-pairs, leaving the name 
$g$-pair for particular correlated pair, i.e. a pair corresponding 
to a specif\/ic isometry $g \in \Gamma$. Similarly, we shall use the
terminology $U$-pairs when referring collectively to uncorrelated
pairs.}
A $u$-pair is a pair $(p,q)$ which is not of the form $(p, gp$) for
any $g \in \Gamma$, that is to say the elements $p$ and $q$ of the
$U$-pairs are {\em not} related (correlated) by any isometry 
$g \in \Gamma$.

Now, if one denotes by $F_g(s_i)$ and $F_u(s_i)$, respectively, the 
probability that the elements of a $g$-pair and of a $u$-pair be 
separated by a distance that lies in $J_i$, the probability $F(s_i)$ 
that a pair in a typical catalog of the ensemble be separated by a 
spatial distance in $J_i$ is given by
\begin{equation} \label{Prob}
F(s_i) = \frac{N_u}{N}\,\,F_u(s_i) 
      + \frac{1}{2}\,\sum_{g \in \widetilde{\Gamma}} 
         \,\, \frac{N_g}{N}\,\, F_g(s_i) \;.
\end{equation}
where  $\widetilde{\Gamma}$ denotes the covering group 
$\Gamma$ without the identity map, and where $N_u$ and $N_g$
denote, respectively, the (total) expected  number of uncorrelated 
pairs and the (total) expected  number of $g$-pairs in a typical 
catalog $\mathcal{C}$ of the ensemble. It should be noticed that
since the pairs of cosmic images are either correlated 
($\Gamma$-pairs) or uncorrelated ($U$-pairs) we must have
\begin{equation} \label{sumNuNg}
N_u + \frac{1}{2}\,\sum_{g \in \widetilde{\Gamma}}\,N_g\,
                = \,N \;. 
\end{equation}

Inserting eq.~(\ref{Prob}) into eq.~(\ref{defepsh}) one 
obtains
\begin{equation} \label{EPSH1}
\Phi_{exp}(s_i) = \frac{N_u}{N}\,\,\Phi^{u}_{exp}(s_i)
      + \frac{1}{2}\,\sum_{g \in \widetilde{\Gamma}} 
         \,\, \frac{N_g}{N}\,\, \Phi^{g}_{exp}(s_i) \;,
\end{equation}
where from~(\ref{defepsh}) we have been led to def\/ine 
the EPSH's corresponding to uncorrelated pairs and 
associated to an isometry $g$, respectively, as 
\begin{eqnarray} 
\Phi^{u}_{exp}(s_i) &=& \frac{1}{\delta s}\, F_u(s_i) =
\frac{1}{N_u}\,\frac{1}{\delta s}\,\, \eta^u_{exp}(s_i)  
                             \label{epshu}\;,\\
\Phi^{g}_{exp}(s_i) &=& \frac{1}{\delta s} \,F_g(s_i) =
\frac{1}{N_g}\,\frac{1}{\delta s}\, \,\eta^g_{exp}(s_i)
                             \label{epshg} \;, 
\end{eqnarray}
with  $N_u = \sum_{s_i} \eta^u_{exp}(s_i)\,$ and 
$\,N_g = \sum_{s_i} \eta^g_{exp}(s_i)\,$.

Similarly for simply connected universe with $N$ pairs of
cosmic images, since all pairs are uncorrelated, 
equation~(\ref{defepsh}) reduces to 
\begin{equation} \label{epshsc}
\Phi^{sc}_{exp}(s_i) = \frac{1}{N}\,\frac{1}{\delta s}\,\,
                            \eta^{sc}_{exp}(s_i)
                       = \frac{1}{\delta s} \,F_{sc}(s_i) \;,
\end{equation}
where $F_{sc}(s_i)$ is the probability that two objects in the 
universe be separated by a distance that lies in $J_i$.

An alternative expression for the EPSH of multiply-connected universe
$\Phi_{exp}(s_i)$ in terms of $\Phi^{sc}_{exp}$ can be obtained. Indeed,
using~(\ref{EPSH1}) and (\ref{sumNuNg}) one easily obtains
\begin{equation}   \label{EPSH4}
\Phi_{exp}(s_i) = \Phi^{sc}_{exp}(s_i)
  + \frac{N_u}{N}\,[\,\Phi^{u}_{exp}(s_i) - \Phi^{sc}_{exp}(s_i)\,] 
  + \frac{1}{2}\,\sum_{g \in \widetilde{\Gamma}}\, 
\frac{N_g}{N} \, [\,\Phi^g_{exp}(s_i) - \Phi^{sc}_{exp}(s_i)\,] \;.
\end{equation}

Using the expression of $N$ in terms of the number of cosmic images $n$ 
from eq.~(\ref{EPSH4}) one finally obtains the following expression 
for what we will def\/ine as the \emph{topological signature} 
of multiply-connected universes, namely
\begin{eqnarray} \label{topsig1}
\varphi^S(s_i) & \equiv & 
(n-1)[\Phi_{exp}(s_i) - \Phi^{sc}_{exp}(s_i)] \nonumber \\
               &=& \varphi^U(s_i) + \varphi^{\Gamma}(s_i) \;,
\end{eqnarray}
where 
\begin{equation}   \label{PhiU}
\varphi^U(s_i) = \nu_u \left[\,\Phi^{u}_{exp}(s_i) 
             - \Phi^{sc}_{exp}(s_i)\,\right] 
\end{equation}
and
\begin{equation}   \label{PhiGamma}
\varphi^{\Gamma}(s_i) = \,\sum_{g \in \widetilde{\Gamma}} 
\nu_g\,[\, \Phi^g_{exp}(s_i) - \Phi^{sc}_{exp} (s_i)\,] \;, 
\end{equation}
and where $\nu_u= 2N_u/n$ and $\nu_g= N_g/n$.

To extract the topological signature $\varphi^S(s_i)$  of multiply-%
connected universes, an important point to bear in mind is that the EPSH 
is essentially a typical PSH from which the statistical noise has been 
withdrawn. Hence we have
\begin{equation}  \label{noise1}
\Phi(s_i) = \Phi_{exp}(s_i) + \rho(s_i) \; ,
\end{equation}
where $\Phi(s_i)$ is a typical PSH constructed from $\mathcal{C}$
and $\rho(s_i)$ represents the statistical f\/luctuation that
arises in the PSH $\Phi(s_i)$. 

\begin{sloppypar}
In practice one can approach the topological signature $\varphi^S(s_i)$
by reducing the statistical f\/luctuations through any suitable 
method to lower the noises $\rho(s_i)\,$.  In computer simulations
the simplest way to accomplish this is to use several comparable 
catalogs to generate a \emph{mean pair separation histogram\/} (MPSH). 
In other words, the use of the MPSH to extract the topological 
signature $\varphi^S(s_i)$ consists in the use of $K$ (say) 
computer-generated comparable  catalogs, with approximately
the same number $n$ of images  and corresponding to the
same manifold $M$, to obtain the mean pair separation histogram
$<\!\Phi(s_i)\!>\,$ (over the $K$ catalogs), and analogously 
to have $<\!\Phi^{sc}(s_i)\!>\,$; and use them as approximations 
for $\Phi_{exp}(s_i)$ and  $\Phi^{sc}_{exp}(s_i)\,$, to construct 
the topological signature $\varphi^S(s_i) \simeq (<\!n\!>-\,1\,)\,
[<\!\Phi(s_i)\!>  - <\!\Phi^{sc}(s_i)\!>]\,$.  
\end{sloppypar}

An improvement of the above procedure to extract the topological 
signature $\varphi^S(s_i)$ comes out for the cases one can 
derive the expression for the PSH's $\,\Phi^{sc}_{exp}(s_i)$
corresponding to the simply connected covering universes. 
The explicit formulae for $\Phi^{sc}_{exp}(s_i)$ corresponding 
to an uniform distribution of objects in the covering universes 
endowed with the Euclidean, elliptic and hyperbolic  geometries 
can indeed be obtained~\cite{BT99,Reboucas00}. 
Thus, for multiply-connected universes with homogeneous distribution 
of objects, which we will be concerned with in the computer simulations, 
the topological signature clearly has the form
$\varphi^S(s_i) \simeq (<\!n\!>-1\,)\,[\,<\!\Phi(s_i)\!> - \,\,
\Phi^{sc}_{exp}(s_i)\,]\,$, where $\Phi^{sc}_{exp}(s_i)\,$ is 
known from the outset.

The f\/irst series of computer-aided simulations concerns a compact 
orientable Euclidean manifold of class ${\mathcal G}_6$ in Wolf's 
classif\/ication~\cite{Wolf}. 
We shall denote this Euclidean cubic manifold by ${\mathcal T}_4$ 
in agreement with the notation used in~\cite{BGRT98}, wherein 
the cubic fundamental polyhedron FP and the pairwise faces 
identif\/ication are shown.
Relative to a coordinate system whose origin coincides with the 
center of the FP, the actions of the generators $\alpha$, $\beta$ 
and $\delta$ on a generic point $p = (x, y, z)$ were shown to be 
described by~\cite{Wolf,Gomero}
\begin{eqnarray}  
\alpha \, p  & = &  (x+L, \,-y, \,-z) \;, \label{isoa}   \\
\beta \, p  & = &  (-x,\, z+L,\, y)   \;, \label{isob} \\
\delta \, p  & = &  (-x,\, z, \, y+L) \; , \label{isod} 
\end{eqnarray}
where $L$ is the edge of the cubic FP. 
In the simulations corresponding to ${\mathcal T}_4$ the center of 
the FP was taken to be the origin of the coordinate system, and to 
coincide with the center of the observed universe $\mathcal{B}_a\,$, 
whose diameter is $2 a = L \, \sqrt{2} \simeq 1.41\,L\,$.
It should be noted that with this ratio for $a/L$ and for
$s \in (0,2a)\,$ one has only the contribution of non-translational 
isometries for the topological signature. 

In the computer simulation we have used a program whose input are 
the number $K$ of catalogs, the radius $a$ of the observed 
universe $\mathcal{B}_a\,$, the number $m$ of subinterval (bins), 
and the number $n_s$ of objects inside the FP (seeds).
The program generates $K$ different catalogs,  starting 
(each) from the same number $n_s$ of homogeneously distributed
seeds inside the FP, and then using the generators $\alpha$, $\beta$ 
and $\delta$ [whose actions are given by~(\ref{isoa})~--~(\ref{isod})]
and their inverses $\alpha^{-1}$, $\beta^{-1}$ 
and $\delta^{-1}$. 
For each  bin $J_i$ of width $\delta s = 2a/m$ it counts the 
normalized number of pairs $\sum \widehat{\eta}_k\,(s)$ for all 
catalogs $k$ from 1 to $K$. 
Finally, it calculates the normalized average numbers of pairs
for all $s_i \in (0,2a)\,$, f\/inding therefore the mean pair 
separations histogram $\,<\!\Phi(s_i)\!>\,$ over $K$ catalogs. 

We have performed simulations corresponding to the manifold 
${\mathcal T}_4$ with $L=1$ and in an observed universe of 
radius $a = \sqrt{2}/2 \simeq 0.71 \,$, $\delta s =0.01$, 
and with dif\/ferent number $n_s$ of seed objects uniformly 
distributed in the FP. Figure~1a is the graph of the topological 
signature $\varphi^{S}(s_i)$ ($\,\simeq (<\!n\!> -\,1\,)\,
[<\!\Phi(s_i)\!>-\,\Phi^{sc}_{exp}(s_i)]\,$) 
and was obtained using the MPSH procedure for $K=16000$ 
catalogs, and  $n_s=15$, which corresponds to an average 
number of images per catalog $<\!n\!>\,\simeq 23$. 
Figure~1b shows a graph of the topological signature for the same
universe and manifold, and was obtained through the MPSH scheme for 
identical number of catalogs, but now the number of seeds was 
taken to be $n_s=100$,  which corresponds to $<\!n\!>\, \simeq 153$ 
images. 
These f\/igures make clear that the topological signature $\varphi(s_i)$
for $n_s=15$ is essentially the same obtained for $n_s=100$, and that 
the plain topological signature arises in simulations where there are 
just a few images for each object. 

We have also performed computer simulations for the specif\/ic elliptic 
3-manifold $S^3/Z_5\,$, whose volume $2\,\pi^2\,R^3/5\,$ is one f\/ifth 
of the volume of the three-sphere $S^3$. A FP (tetrahedron) together 
with the pairwise faces identif\/ications is given by Weeks~\cite{Weeks85}. 
We have taken as the observed universe the whole covering space $S^3$, i.e.
a solid sphere with radius $a=\pi$ ($\,R=1\,$ in the corresponding
RW metric, and the edge of tetrahedron $L \simeq 1.82$). 
Thus all catalogs in our simulations for this manifold have the 
same number of images.  
Figure~2 shows the graph of the topological signature $\varphi^S(s_i)$ 
for this multiply-connected universe, for $m=180$, $n=100$ images 
($n_s=20$), $K=3000$ catalogs. 

We have finally performed computer simulations for the specif\/ic compact
hyperbolic 3-manifold known as Seifert-Weber dodecahedral space, 
which is obtained by identifying or glueing the opposite pentagonal 
faces of a dodecahedron after a rotation of $3\,\pi/5\,$.  
Figure~3 shows the graph of the topological signature $\varphi^S(s_i)$
for this hyperbolic space  where the center of the dodecahedron was 
taken to coincide with the center of the observed universe 
$\mathcal{B}_a\,$, whose diameter is $2a \simeq 2.88$. The length $L$ of 
the edges of the pentagonal faces and the height $H$ of the dodecahedron 
are $L=H \simeq 1.99\,$, where the lengths are measured with the 
hyperbolic RW geometry with $R=1$. 
We have taken $m=100\,$ bins, $n_s=10$ seeds ($<\!n\!>\, \simeq 18$), 
$K=16000$ catalogs, and used the exact expression for 
$\Phi^{sc}_{exp}(s_i)$.

To close this work it is worthwhile mentioning that the ultimate 
step in most of such statistical approaches to extract the 
topological signature is the comparison of the graphs (signature)
obtained from simulated catalogs against similar graphs generated 
{}from real catalogs. To do so one clearly has to have the simulated 
patterns of the topological signatures of the universes, which can
be achieved by the method we have proposed in this article.

\vspace{1cm}
\section*{Captions for the figures}
\begin{description}

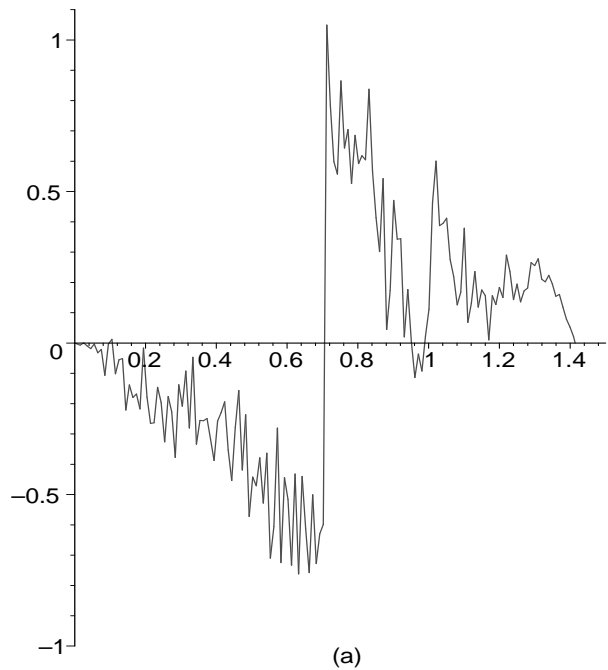
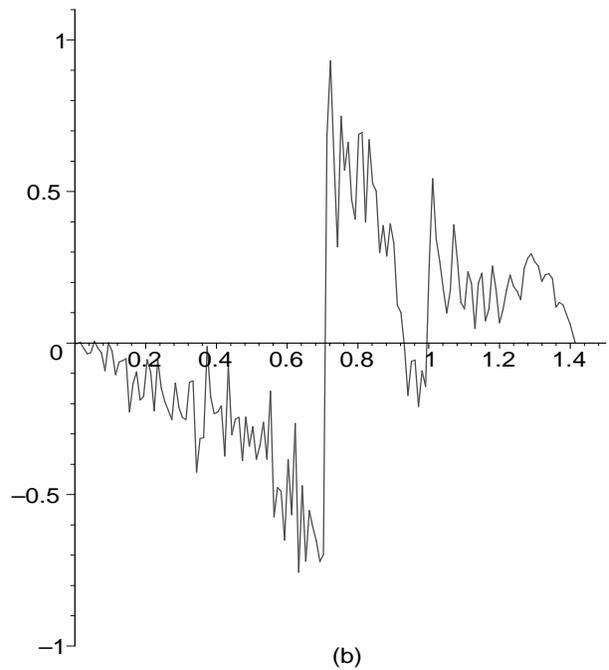
\item[Figure 1] 
The topological signature $\varphi^S(s_i)$ of an Euclidean multiply-%
connected universe of diameter $2a \simeq 1.41$, with underlying 
topology of ${\mathcal T}_4$ with edge $L=1$.
The horizontal axis gives the pair separation $s$ while the
vertical axis furnishes the normalized number of pairs.
In (a) the number of seeds is $n_s=15$ and corresponds to an 
average number of images per catalog $<\!n\!> \,\,\simeq 23\,$.
In (b) the number of seeds is $n_s=100$ and corresponds to an 
average number of images per catalog $<\!n\!>\,\, \simeq 153\,$.
In both cases one arrives at essentially the same topological
signature.

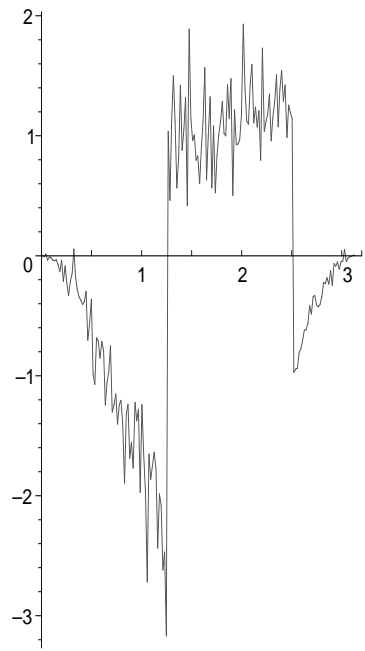
\item[Figure 2]
The topological signature $\varphi^S(s_i)$ of an elliptic multiply-%
connected universe with diameter $2a=\pi$ and topology $S^3/Z_5$.
The edge of the tetrahedron (FP) is $L \simeq 1.82$. The observed 
universe is whole unitary sphere $S^3$.
The horizontal axis gives the pair separation $s$ while the
vertical axis gives the normalized number of pairs.
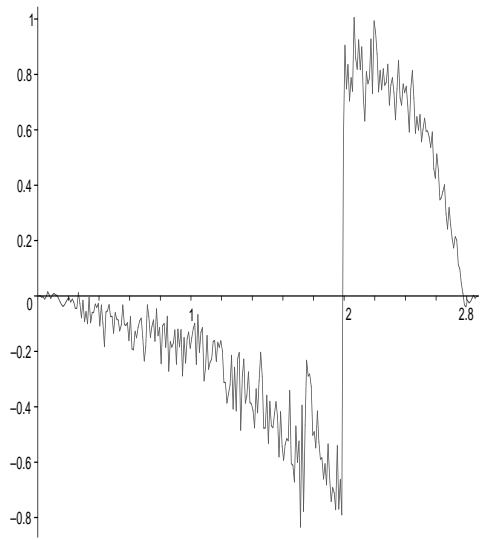
\item[Figure 3]
The topological signature $\varphi^S(s_i)$ for a hyperbolic multiply-%
connected universe with diameter $2a \simeq 2.88$ whose underlying 
topology is the Seifert-Weber dodecahedral space with edge and height 
$L=H \simeq 1.99\,$. 
The horizontal axis gives the pair separation $s$ while the
vertical axis provides the normalized number of pairs.
\end{description}

\end{document}